\documentclass[
 superscriptaddress,
 amsmath,amssymb,
 aps,
 floatfix
]{revtex4-2}

\usepackage{amsmath}
\usepackage{amssymb}
\usepackage{amsthm}
\usepackage{bm}
\usepackage{graphicx}
\usepackage{hyperref}
\usepackage[capitalise]{cleveref}
\usepackage[a-1b]{pdfx}

\newtheorem{conjecture}{Conjecture}

\begin{document}

\title{On a Finite Population Variation of the Fisher-KPP Equation}

\author{Christopher Griffin}
\email{griffinch@psu.edu}
\affiliation{
	Applied Research Laboratory,
    The Pennsylvania State University,
    University Park, PA 16802
    }

\begin{abstract}
In this paper, we formulate a finite population variation of the Fisher-KPP equation using the fact that the reaction term can be generated from the replicator dynamic using a two-player two-strategy skew-symmetric game. We use prior results from Ablowitz and Zeppetella to show that the resulting system of partial differential equations admits a travelling wave solution, and that there are closed form solutions for this travelling wave. Interestingly, the closed form solution is constructed from a sign-reversal of the known closed form solution of the classic Fisher equation. We also construct a closed form solution approximation for the corresponding equilibrium problem on a finite interval with Dirichlet and Neumann boundary conditions. Two conjectures on these corresponding equilibrium problems are presented and analysed numerically. 
\end{abstract}

\maketitle

\section{Introduction}
Fisher \cite{F37} proposed the following equation
\begin{equation}
\frac{\partial u}{\partial t} = r u (1-u) + \frac{\partial^2 u}{\partial x^2}
\label{eqn:Fisher}
\end{equation} 
as a model of the propagation of a mutant genes. At (approximately) the same time, Kolmogorov, Petrovsky and Piskunov analysed a more general version of \cref{eqn:Fisher} \cite{KPP37} with $ru(1-u)$ replaced by a generic function $F(u)$ having prerequisite smoothness and end point properties. Throughout this paper, we refer to \cref{eqn:Fisher} as the Fisher-KPP equation. This equation is known to appear in simple models of susceptible-infected (SI) spatial epidemics with drift \cite{K04,SRC15,BBP21}. It also emerges naturally (often as a simplification) in branching processes \cite{M12}, for modelling the spread of invasive species \cite{NP04} and in flame propagation and combustion \cite{ZR02} among other areas.

Solutions (especially travelling wave solutions) to the Fisher-KPP equation have been studied by several authors. Fisher himself studied the existence of wave solutions \cite{F37} as did Kolmogorov, Petrovsky and Piskunov \cite{KPP37}. Kametka \cite{K76} and Uchiyama \cite{U77} studied the asymptotic formation of travelling wave solutions. Later, Newman \cite{N80} studied exact solutions. Relevant and related to the brief analysis presented here, Weinberger analysed a discrete form of the Fisher-KPP equation \cite{W82}. However, the most relevant result for this work is from Ablowitz and Zeppetella \cite{AZ79} who provide an explicit solution for the Fisher-KPP equation. 

The objective of this paper is to construct a system of partial differential equations modelling both the propagation of a dominant species (allele, infective, competitor etc.) in the presence of a (finite) population experiencing its own spatial evolution. We do so by using a recent formulation by Griffin, Mummah and DeForest for a finite population spatial replicator equation \cite{GMD21}. This work extends earlier work by (among others) Vickers \cite{V91} who the studied the spatial replicator with an (assumed) infinite population and work by Durrett and Levin \cite{DL94} who compared discrete and spatial population dynamics to continuous population dynamics. We show that this system of partial differential equations admits all the travelling wave solutions of the Fisher-KPP equation, except with directions reversed. We use this fact and the results from \cite{AZ79} to derive an explicit example and interpret it in a physical context. We then study the corresponding equilibrium problem for this equation system on a finite interval. We derive a closed form approximation for the Dirichlet problem and use it to construct a conjecture  on solution behaviours as the spatial gradient of the population becomes large. A simpler problem with Neumann boundary conditions is also briefly considered. 

The remainder of this paper is organized as follows: We derive the model to be studied in Section \ref{sec:ModelDeriv}. In \cref{sec:TW} we show that travelling wave solutions exist and construct an explicit example. We study the equilibrium problem for the equation system in \cref{sec:Asymptotic}, constructing a closed form approximation for the problem with Dirichlet boundary conditions as well as two conjectures (to be proven in future work). Conclusions and future directions are presented in \cref{sec:Conclusion}.

\section{Model Derivation}\label{sec:ModelDeriv}
The classic Fisher-KPP equation, \cref{eqn:Fisher}, can be derived as an example of the one-dimensional infinite population spatial replicator equation 
\begin{equation}
\frac{\partial u_i}{\partial t} = u_i\left(\mathbf{e}_i - \mathbf{u}\right)^T\mathbf{A}\mathbf{u} + k\frac{\partial^2 u_i}{\partial x^2},
\label{eqn:SpatialReplicator}
\end{equation}
with skew-symmetric payoff matrix,
\begin{equation}
\mathbf{A} = \begin{bmatrix}0 & -r\\r & 0\end{bmatrix}.
\label{eqn:Payoff}
\end{equation}
Here $\mathbf{u} = \langle{u_1,u_2}\rangle$ and $k$ is a diffusion constant. With this payoff matrix $\mathbf{u}^T\mathbf{A}\mathbf{u} = 0$. Focusing on the equation for $u_2$,  we can write:
\begin{equation}
\frac{\partial u_2}{\partial t} = r u_2u_1 + k\frac{\partial^2 u_2}{\partial x^2}.
\end{equation}
From \cite{GMD21} we know that for all $(x,t)$ we have $u_1 + u_2 = 1$. Therefore, when $i = 2$, we recover the Fisher-KPP equation
\begin{equation}
\frac{\partial u_2}{\partial t} = r u_2(1-u_2) + k\frac{\partial^2 u_2}{\partial x^2},
\end{equation}
which can simply be written as \cref{eqn:Fisher} with no subscripts and assuming $k=1$.

In \cite{GMD21} is it shown that for finite populations, the spatial replicator is given by the system of equations
\begin{equation*}
\begin{aligned}
&\forall r \left\{
\begin{aligned}
\frac{\partial u_r}{\partial t} =  u_r\cdot\left(\mathbf{e}_r^T\mathbf{A}\mathbf{u} -\mathbf{u}^T\mathbf{A}\mathbf{u}\right)+ 
k\left(\frac{2}{M}\nabla M \cdot \nabla u_r +  \Delta u_r\right)
\end{aligned}\right.\\
&\frac{\partial M}{\partial t} = M\mathbf{u}^T\mathbf{A}\mathbf{u} + k\Delta M.
\end{aligned}
\end{equation*}
Here $M(x,t) > 0$ is the total population of all species at $(x,t)$. If $M \to \infty$ and $\partial_x M$ is bounded or $\partial_x M = 0$, and we restrict to one dimension, then this equation is identical to \cref{eqn:SpatialReplicator}. Using the payoff matrix defined in \cref{eqn:Payoff} and restricting to one dimension, we obtain the one dimensional finite population Fisher-KPP system
\begin{equation}
\left\{
\begin{aligned}
\frac{\partial u}{\partial t} &= r u(1-u) + \frac{2k}{M}\frac{\partial M}{\partial x}\frac{\partial u}{\partial x} + k\frac{\partial^2 u}{\partial x^2}\\
\frac{\partial M}{\partial t} &= k\frac{\partial^2 M}{\partial x^2}.
\end{aligned}
\right.
\end{equation}
The population equation is simply a diffusion equation and so (without boundary or initial conditions) we are free to choose any suitable solution. We can use this to find a travelling wave solution to the system of equations, and consequently a closed form solution.

\section{Travelling Wave Solution}\label{sec:TW}
The population (diffusion) equation has a travelling wave solution given by
\begin{equation}
M(x,t) = \pm\frac{A}{c}\exp\left[\pm c(x\pm kct)\right] + B,
\label{eqn:PopulationTravelingWave}
\end{equation}
where $A$ and $B$ are arbitrary constants. Assume $B = 0$. Then
\begin{equation}
\frac{1}{M}\frac{\partial M}{\partial x} = \pm c.
\end{equation}
Let $k = 1$ and let $z = x\pm ct$. A travelling wave solution for $u$ can be found by solving
\begin{equation}
\pm cu' = r u(1-u) \pm 2c u' + u'',
\end{equation}
assuming $\lim_{x\to\infty} u(x) = 0$ and $\lim_{x\to -\infty}u(x) = 1$, following \cite{AZ79}. Simplifying yields
\begin{equation}
r u(1-u) \pm c u' + u'' = 0.
\label{eqn:FiniteTravellingWave}
\end{equation}
The travelling wave differential equation for \cref{eqn:Fisher} (the Fisher-KPP equation) is 
\begin{equation}
r u (1-u) \mp cu' + u'' = 0.
\end{equation}
This is identical to \cref{eqn:FiniteTravellingWave} except for the sign of the $cu'$ term, which is reversed. Therefore, the finite and infinite population equations must share travelling wave solutions but with their directions of travel reversed, assuming the total population is given by \cref{eqn:PopulationTravelingWave}.

\begin{figure}[htbp]
\centering
\includegraphics[width=0.98\textwidth]{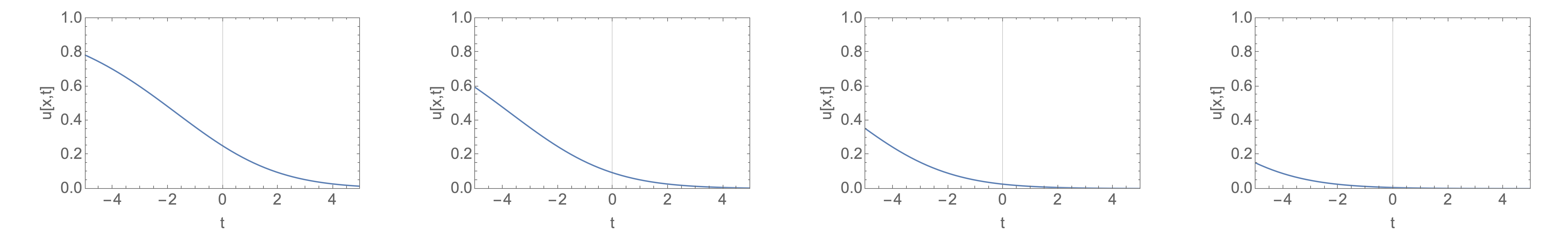}\\
\includegraphics[width=0.98\textwidth]{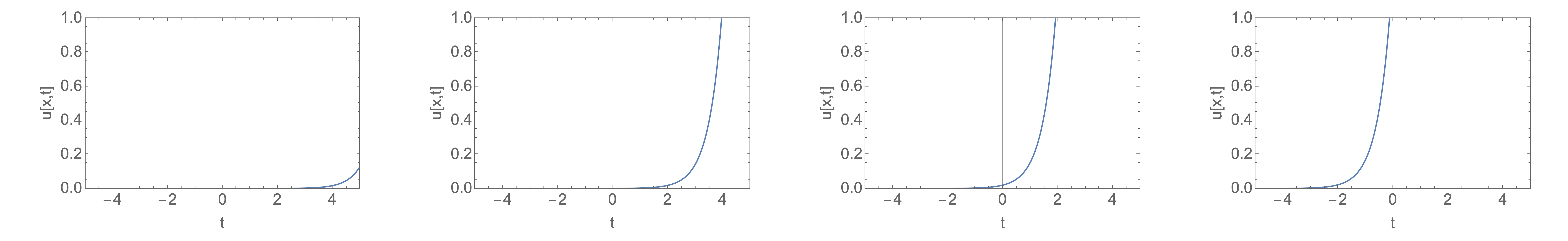}
\caption{(Top) The travelling wave solution $u(x,t)$ given by \cref{eqn:ClosedSolution} assuming $c > 0$ given by \cref{eqn:WaveSpeed} and $k = 1$. (Bottom) The travelling wave solution for the population $M(x,t)$ with $k = 1$ and $c > 0$ given by \cref{eqn:WaveSpeed}. For scaling, we set $A = 10^{-5}$.} In both figures, slices are at $t = \in \{0,1,2,3\}$.
\label{fig:Solution}
\end{figure}

If we rescale so that $r = 1$, then the (astounding) results of Ablowitz and Zeppetella \cite{AZ79} imply, for the special wave speed of:
\begin{equation}
c = \pm \frac{5}{\sqrt{6}},
\label{eqn:WaveSpeed}
\end{equation}
we have the fully closed form solution:
\begin{equation}
u(z) = \left[1 + C\exp\left(\pm \frac{z}{\sqrt{6}}\right)\right]^{-2},
\label{eqn:ClosedSolution}
\end{equation}
where $C$ is another arbitrary constant. We note the solution differs from that in \cite{AZ79} exactly in the sign of $z$, as a result of the introduction of the finite population term. As we require $M(x,t) > 0$, it follows that there is only one acceptable solution when $c > 0$. This is illustrated in \cref{fig:Solution}.
The result is biologically interesting because Ablowitz and Zeppetella's construction assumes that $u(-\infty) = 1$ and $u(\infty) = 0$. We can think of the payoff matrix given in \cref{eqn:Payoff} as describing a prisoner's dilemma game, with strategy two being the dominant one (defect) and strategy one (cooperate) being dominated. (This is equivalent to an SI model in which strategy two is the infected state.) 
\cref{fig:Solution} illustrates a solution describing an infinite wave of cooperators who overwhelm the defector population as they move from right to left. This is somewhat contrary to our intuition, which suggests the defectors should invade the travelling population wave.

\section{Asymptotic Behaviour in Finite Regions}\label{sec:Asymptotic}
\subsection{Dirichlet Boundary Conditions}
\begin{figure}[htbp]
\centering
\includegraphics[width=0.3\textwidth]{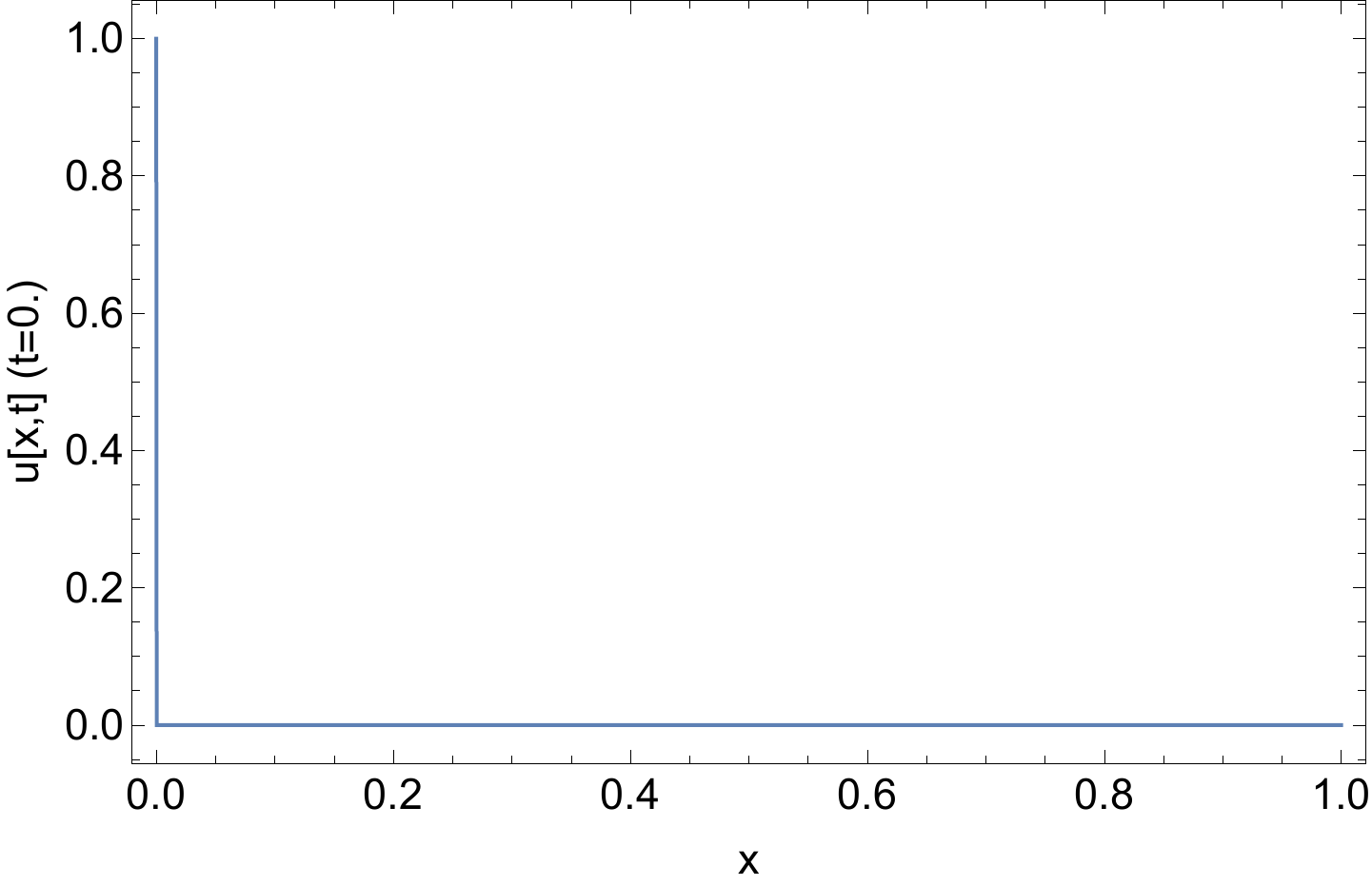}\quad
\includegraphics[width=0.3\textwidth]{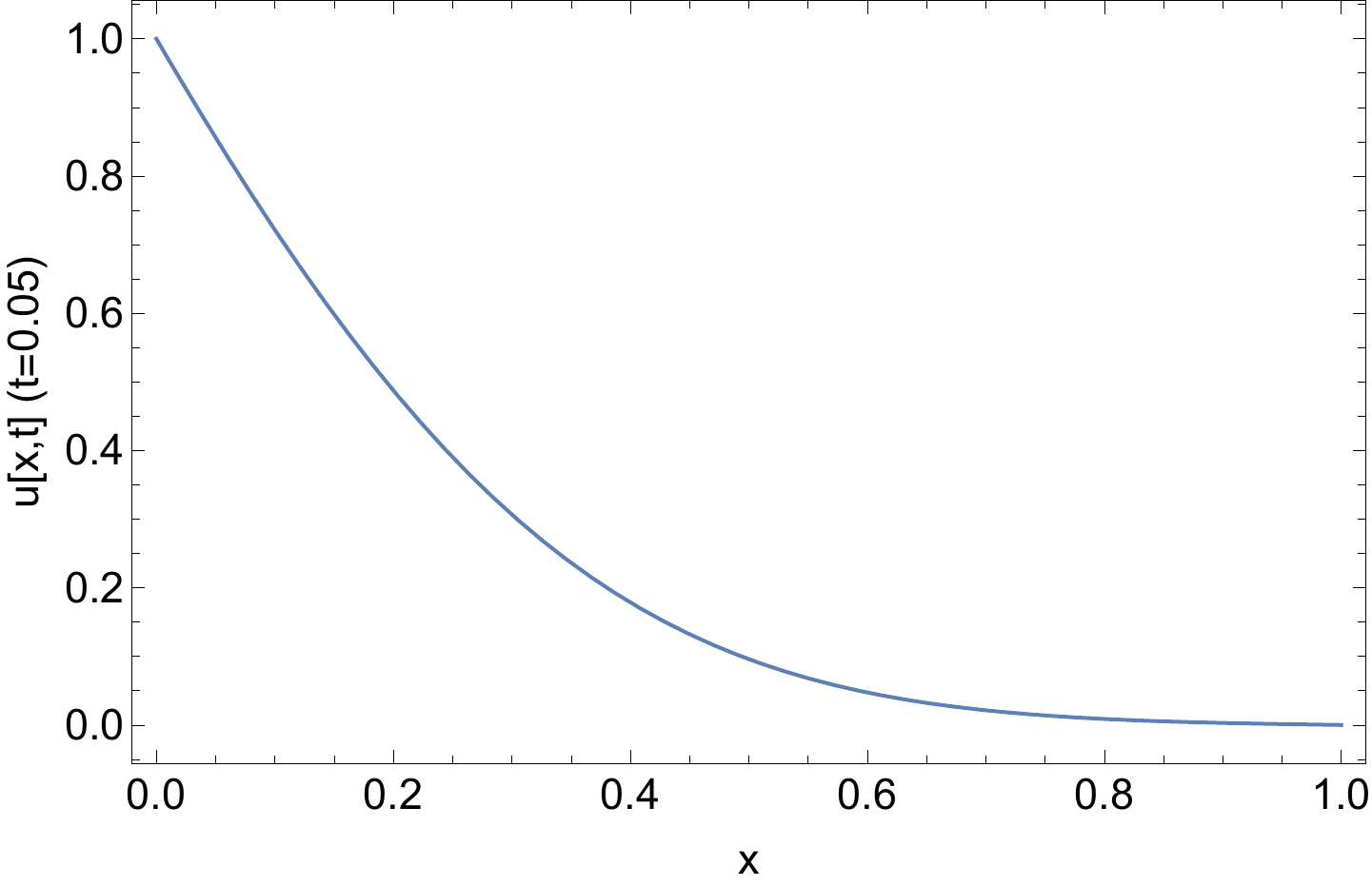}\quad
\includegraphics[width=0.3\textwidth]{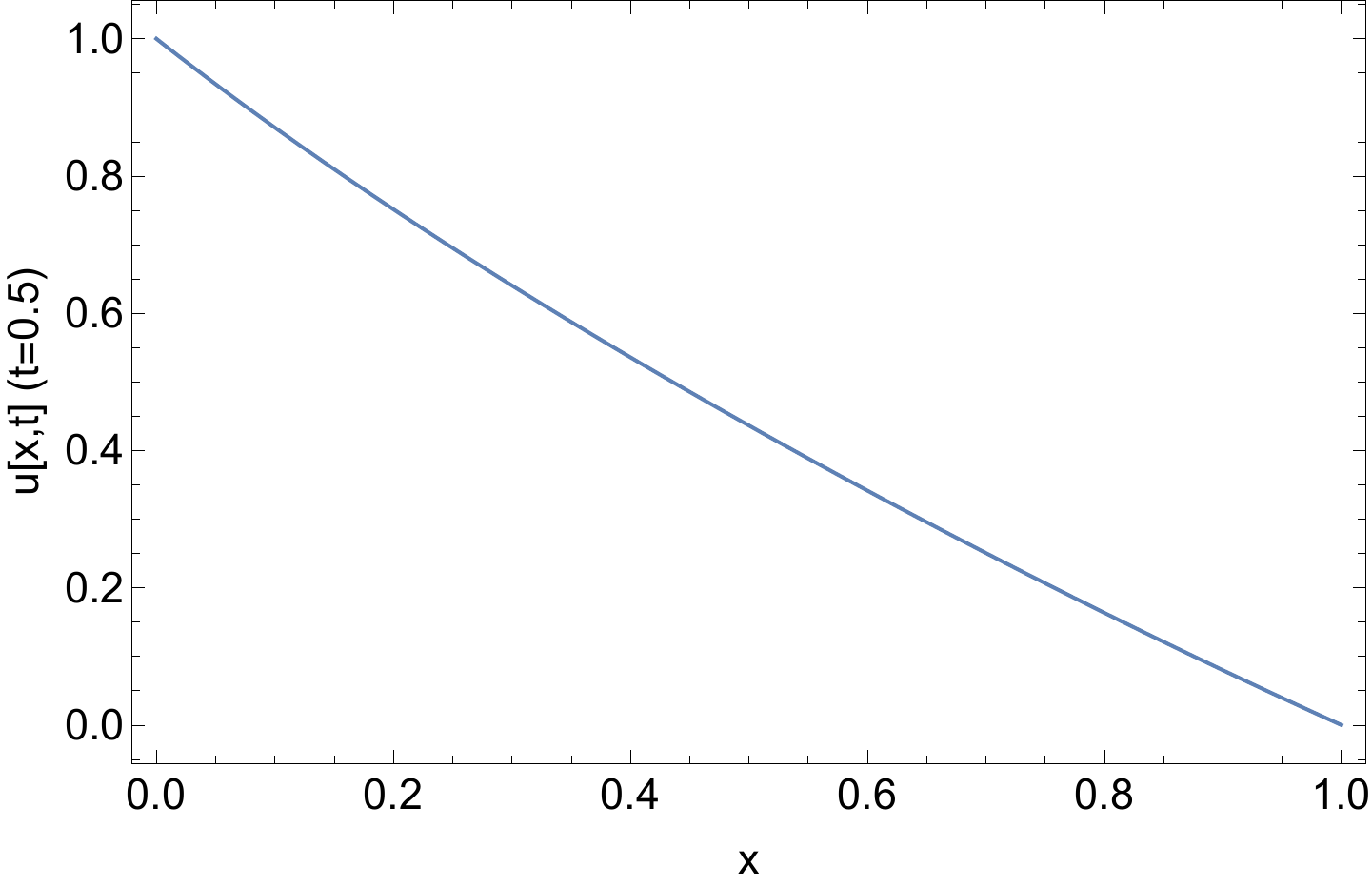}\\
\includegraphics[width=0.3\textwidth]{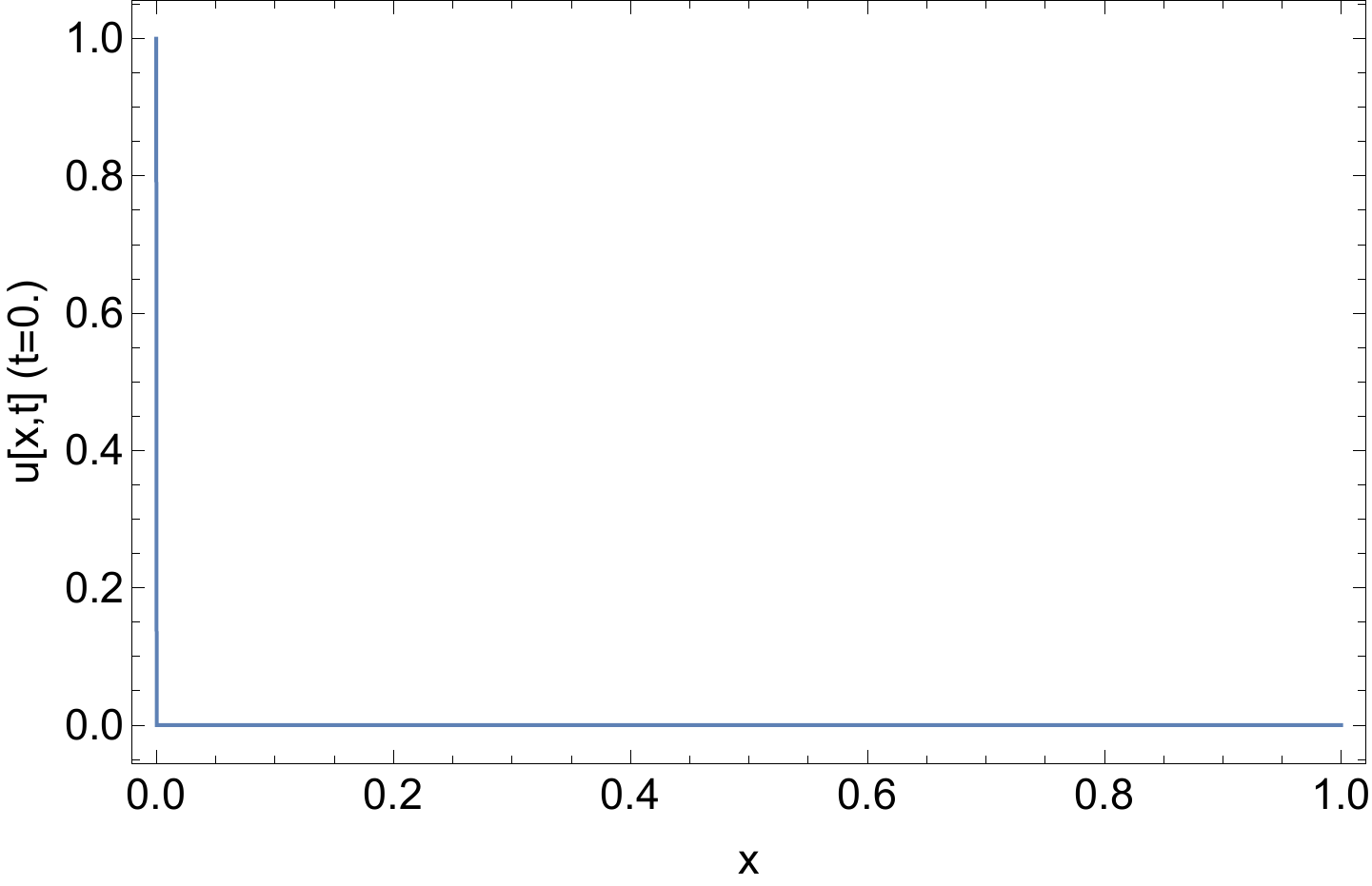}\quad
\includegraphics[width=0.3\textwidth]{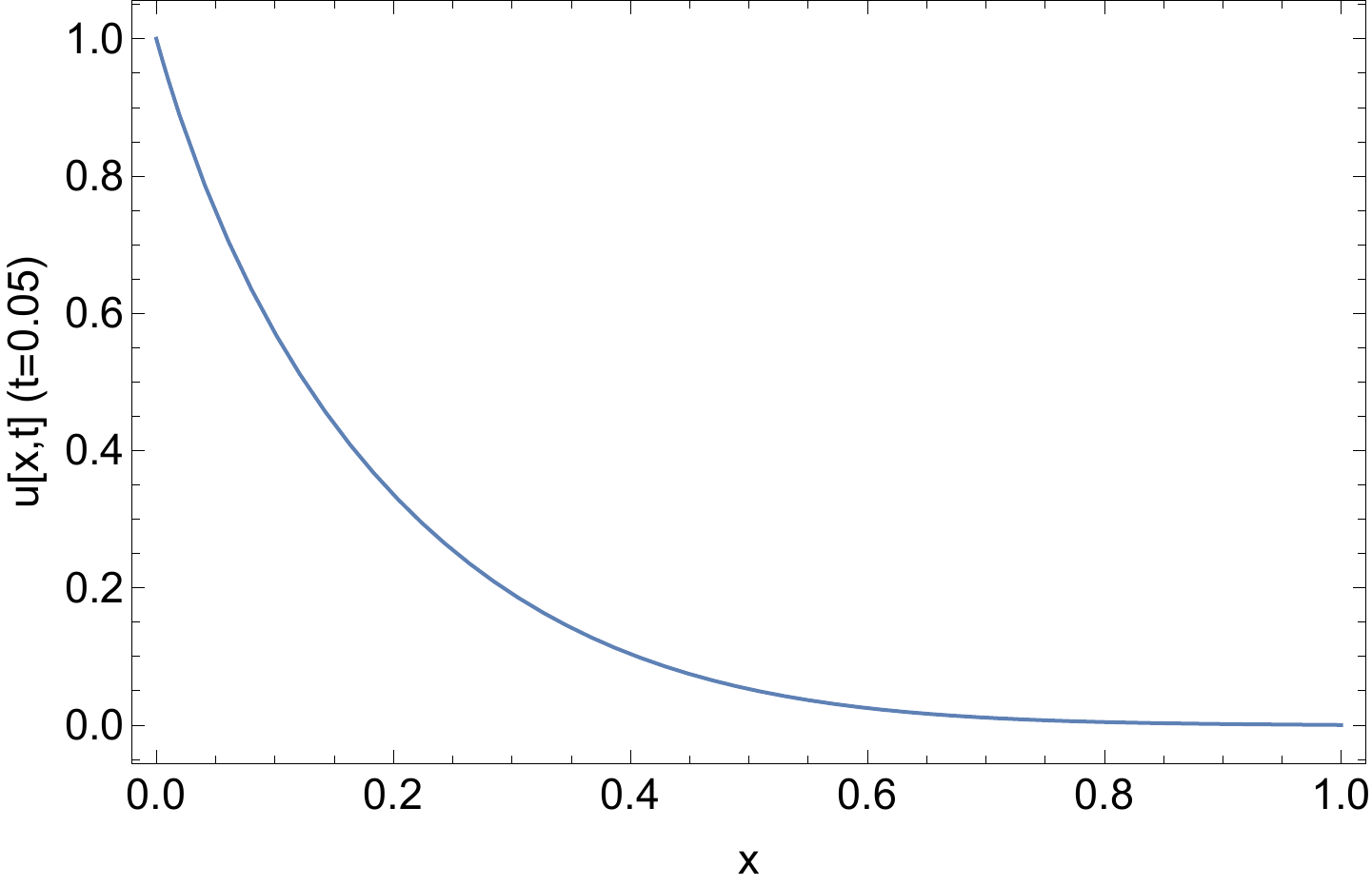}\quad
\includegraphics[width=0.3\textwidth]{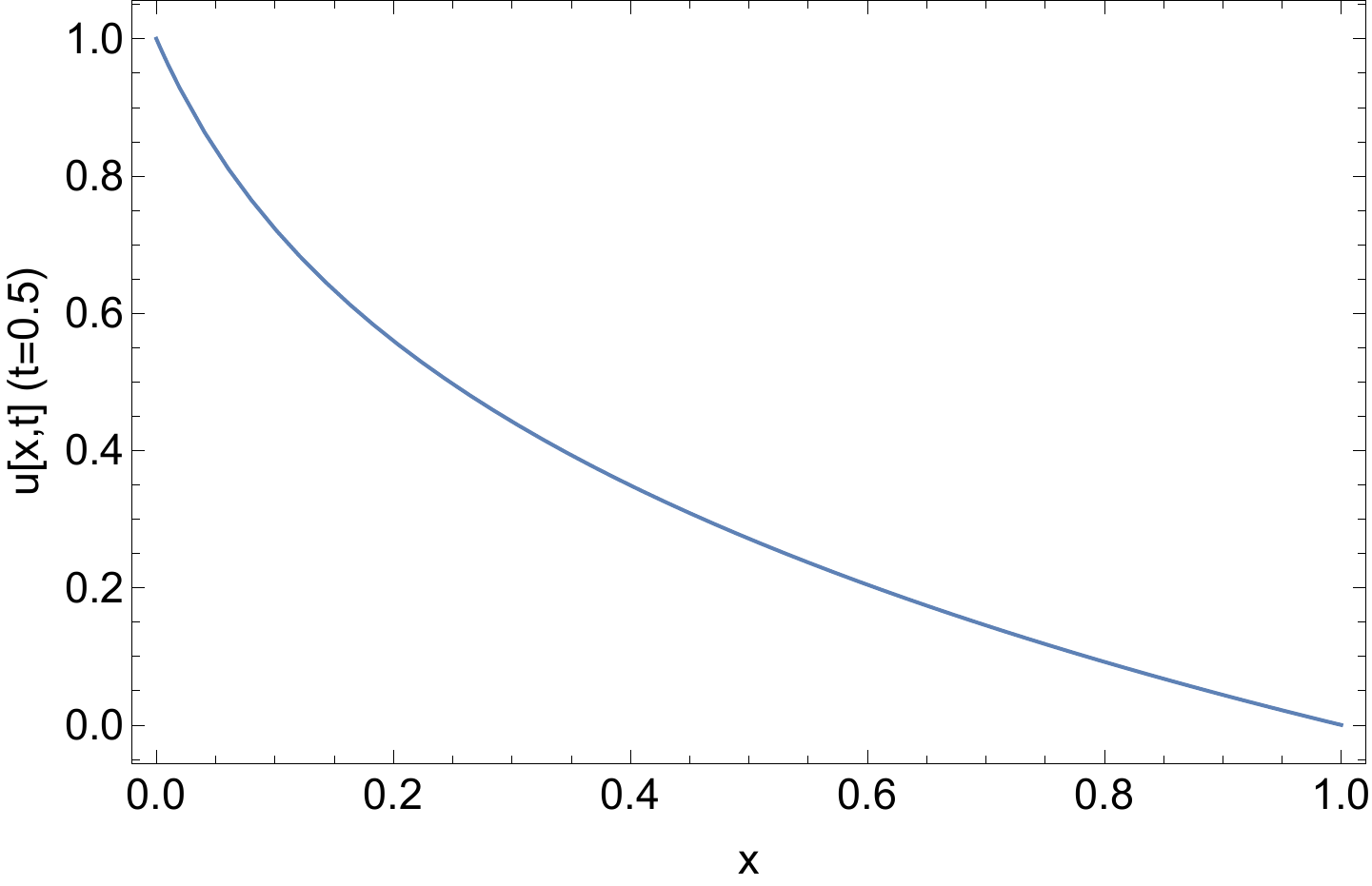}
\caption{(Top Row) Numerical solution for $u(x,t)$ assuming an invasion from the left population described by $M(x) = 1 + x$.
(Bottom Row) Numerical solution for $u(x,t)$ assuming an invasion from the left population described by $M(x) = 1 + 10x$. Both solutions have $k = r = 1$.}
\label{fig:FiniteRegionSpaceTime}
\end{figure}

\begin{figure}[htbp]
\centering
\includegraphics[width=0.65\textwidth]{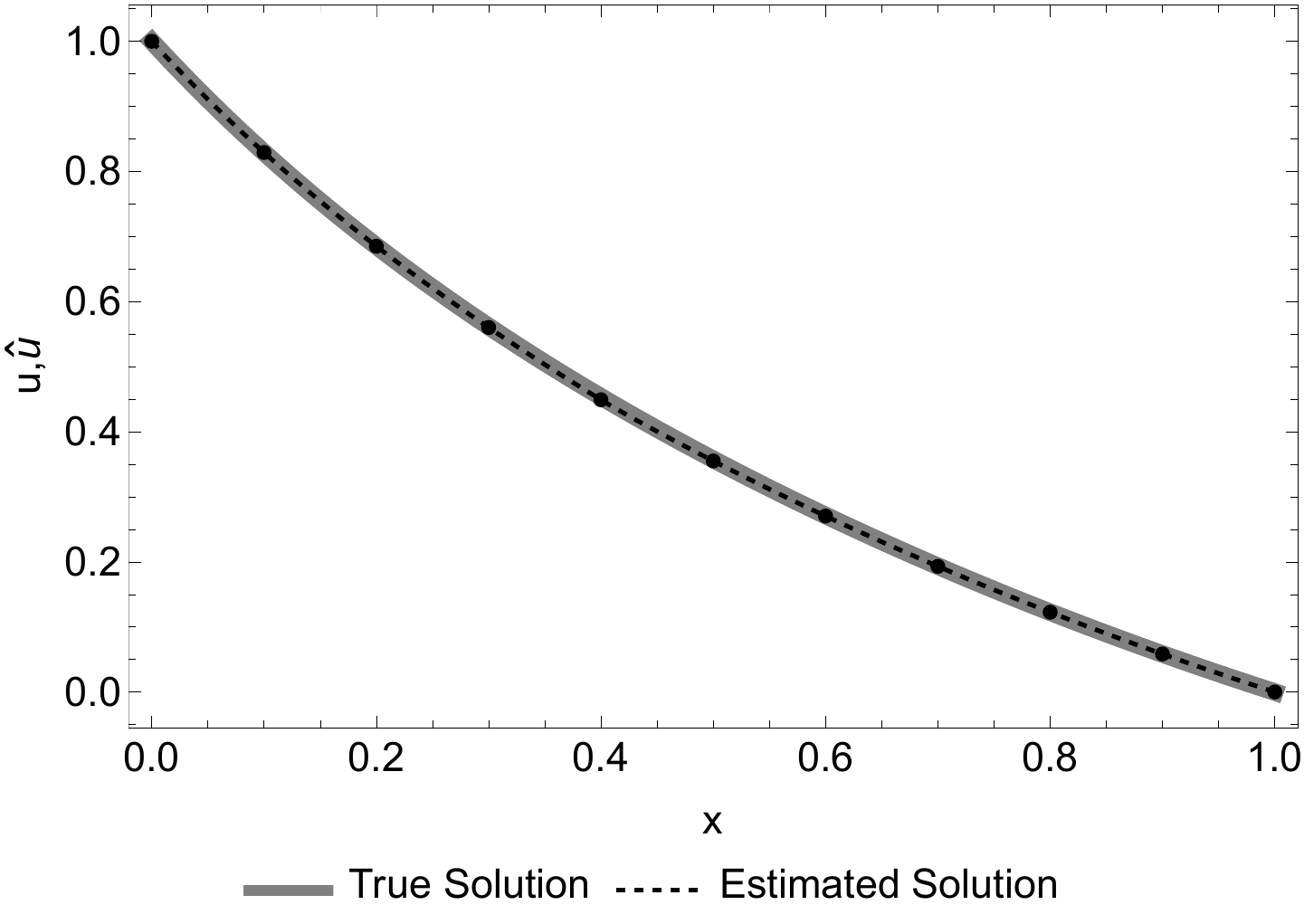}
\caption{A comparison of $\hat{u}$ and $u$ showing goodness of fit of the approximate solution. The computed parameters are $q=0.413$, $r_L=-1.878$ and $r_R=-0.556$.}
\label{fig:uhat}
\end{figure}

\begin{figure}[htbp]
\centering
\includegraphics[width=0.6\textwidth]{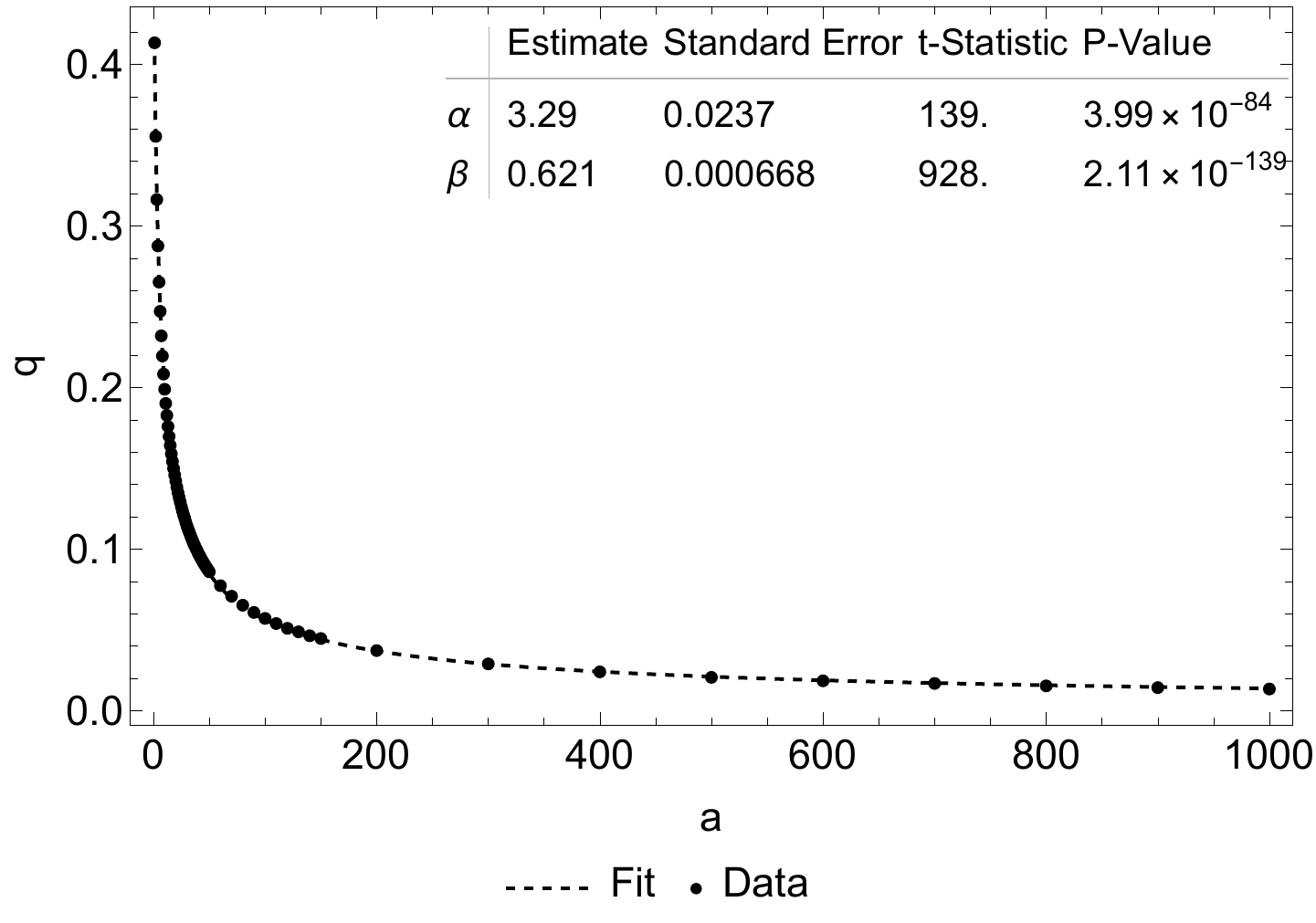}\\
\includegraphics[width=0.6\textwidth]{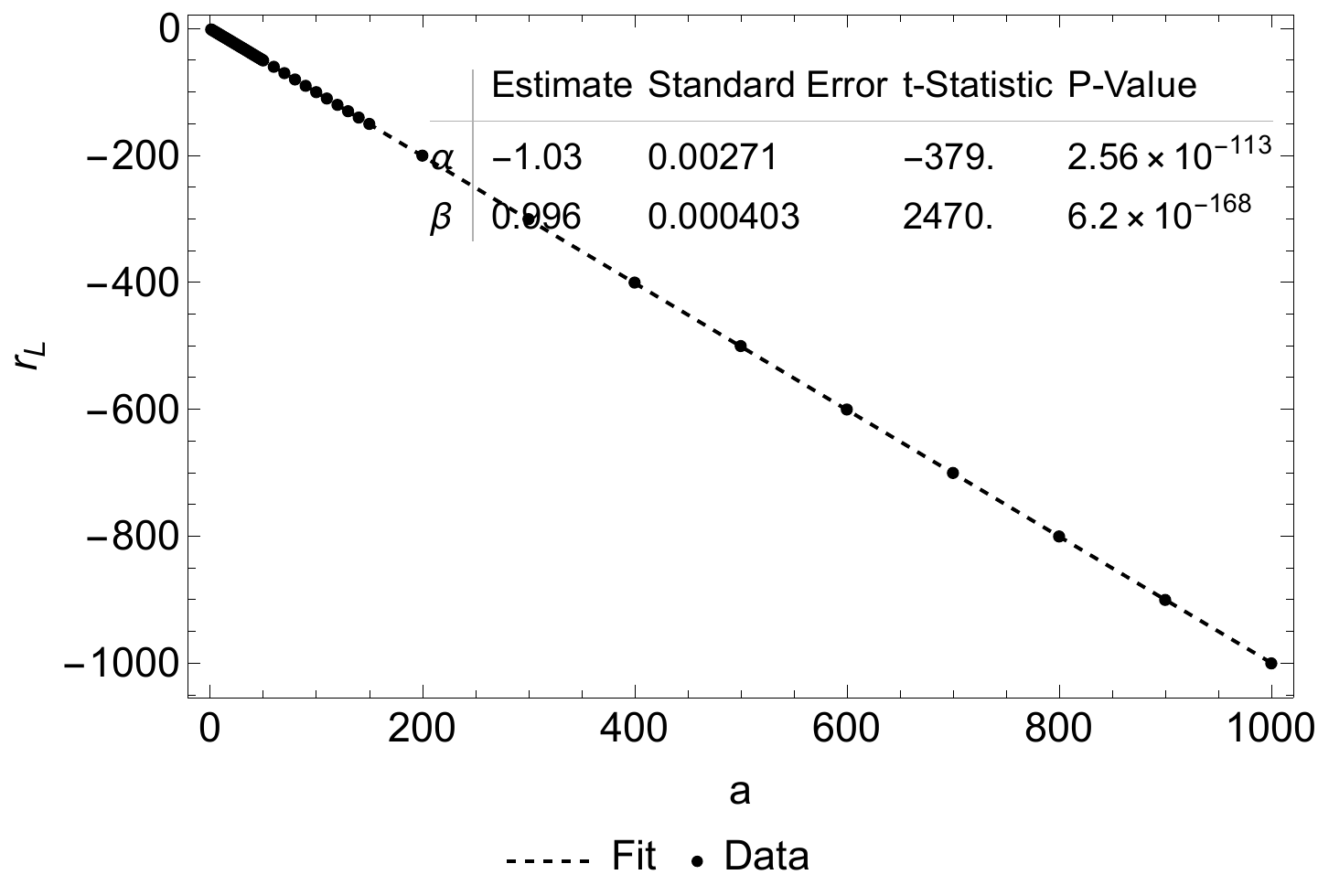}\\
\includegraphics[width=0.6\textwidth]{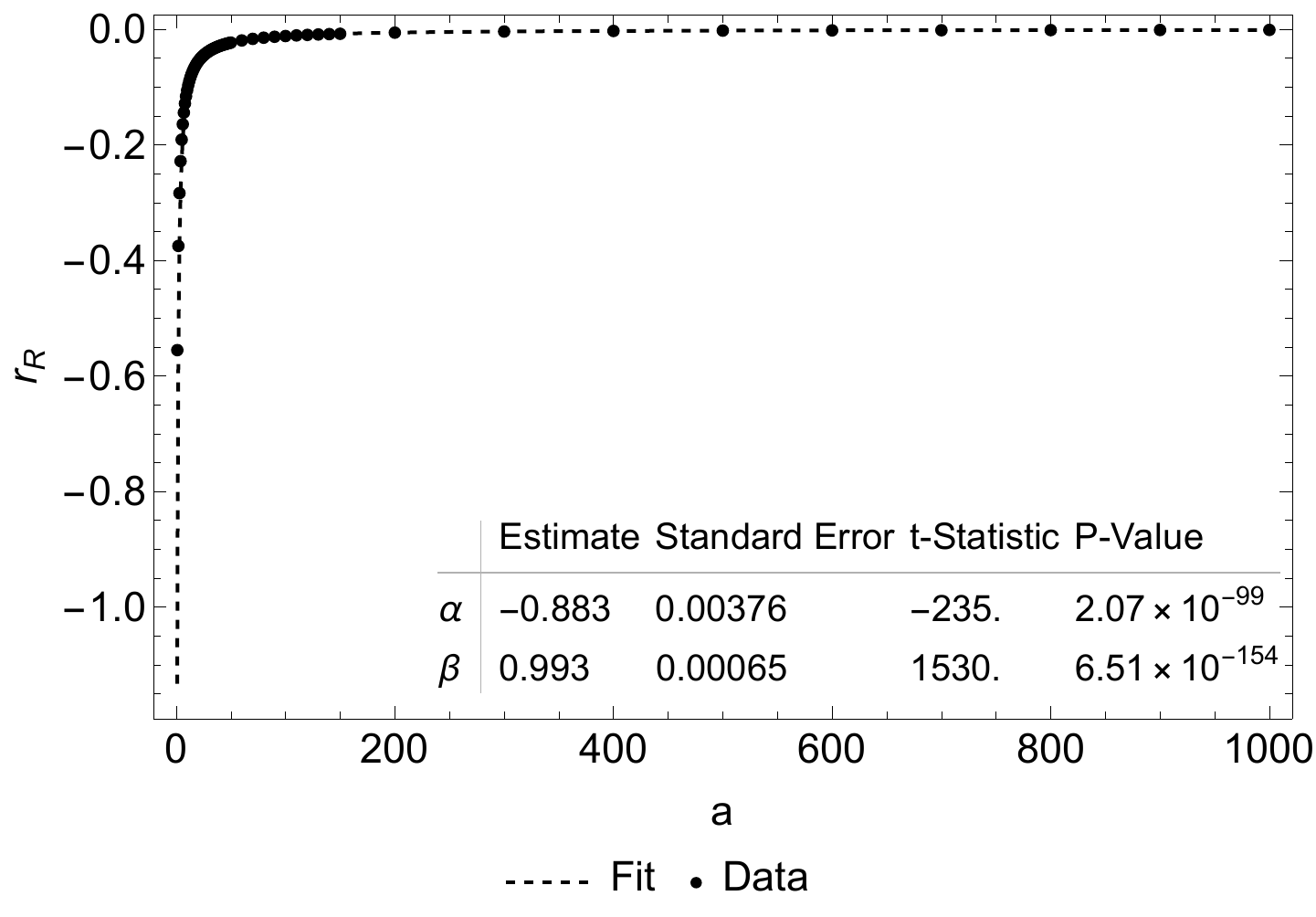}
\caption{(Top) Data and corresponding fit for $q(a)$. (Center) Data and corresponding fit for $r_L(a)$. Despite the appearance of linearity, there is a slight curvature. (Bottom) Data and corresponding fit for fit of $r_R(a)$.}
\label{fig:FitPlots}
\end{figure}

\begin{figure}[htbp]
\centering
\includegraphics[width=0.48\textwidth]{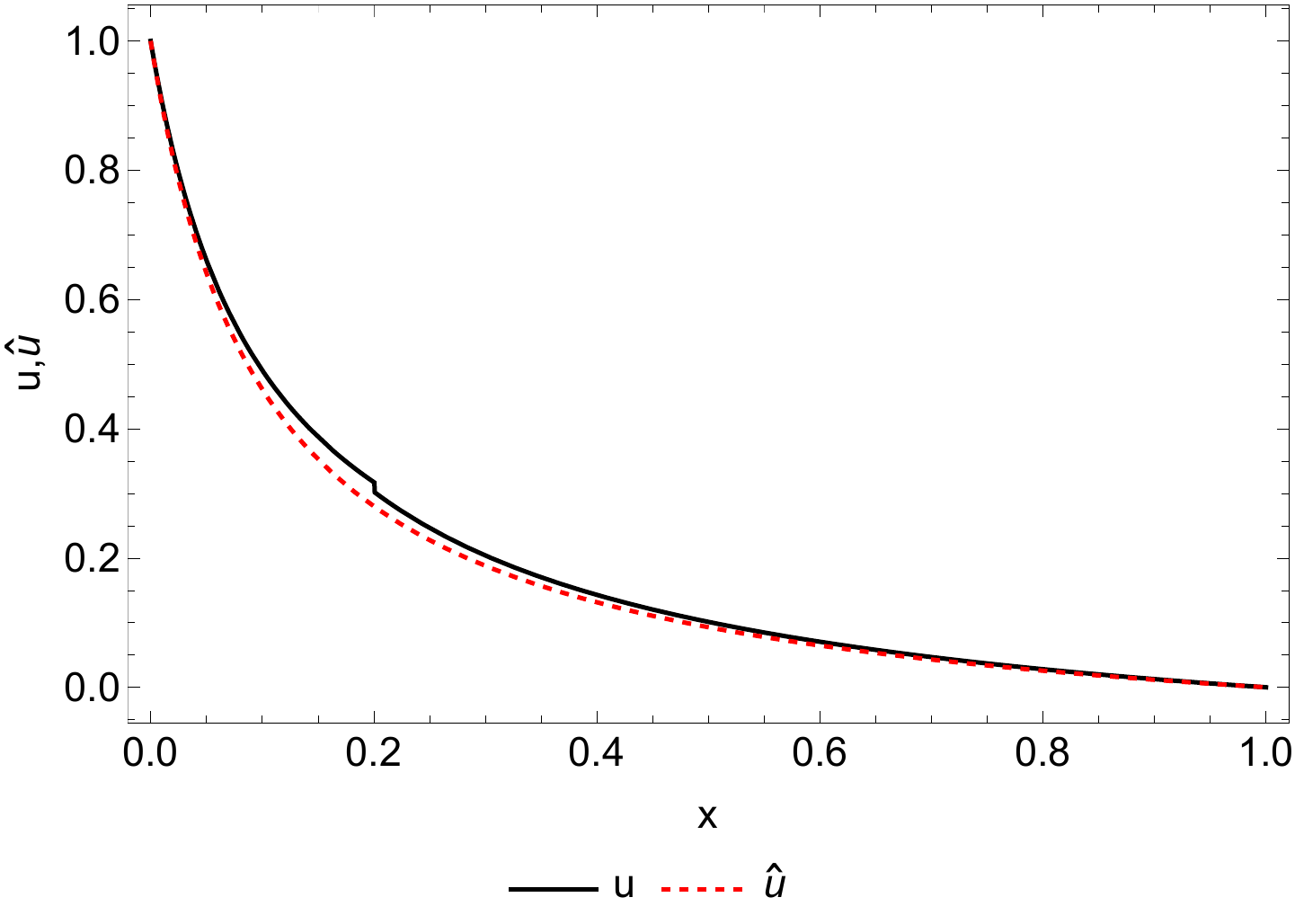}\quad
\includegraphics[width=0.48\textwidth]{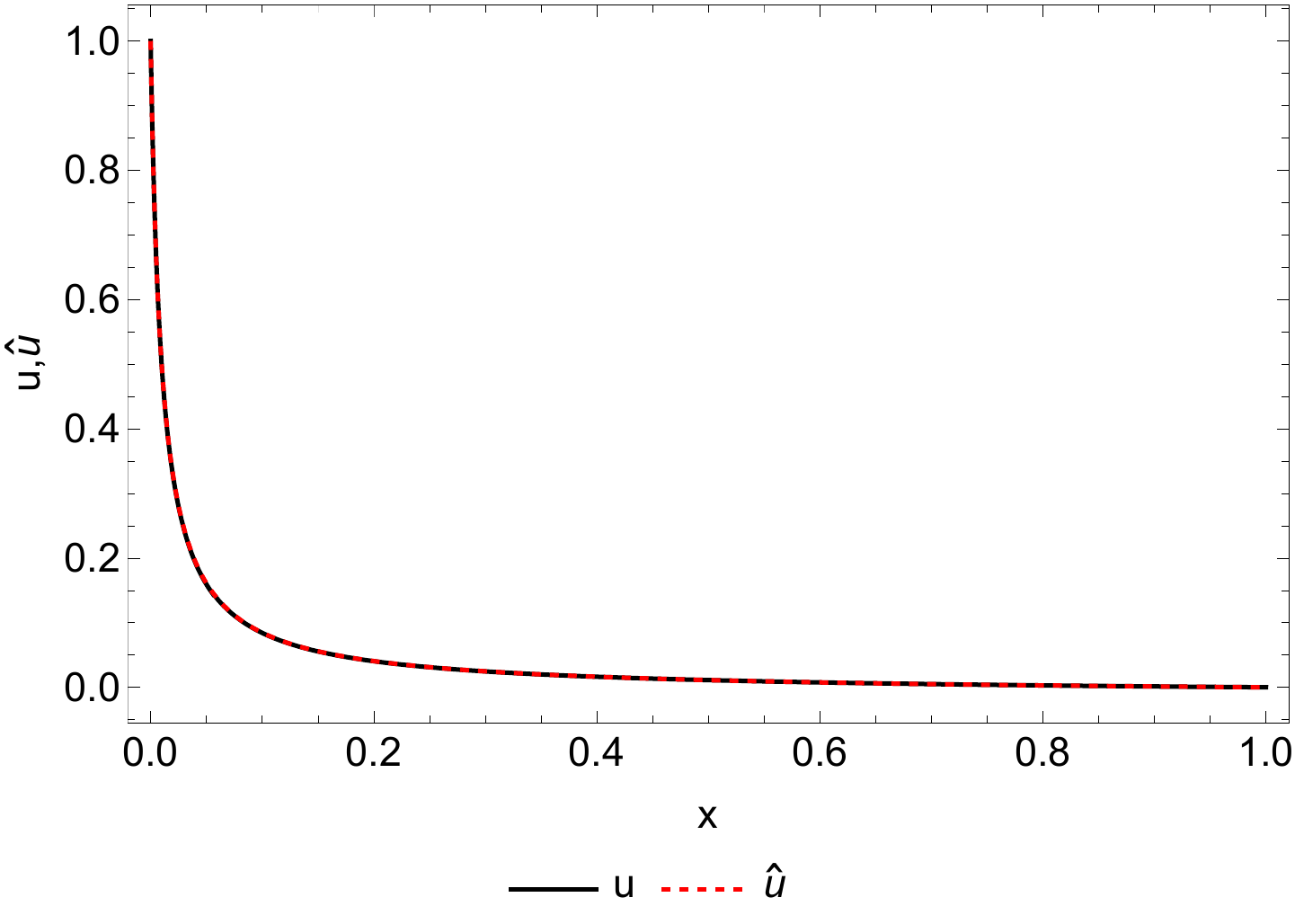}\quad
\caption{(Left) The approximation $\hat{u}$ and $u$ for $a = 10$ shows a reasonable fit with a minor discontinuity. (Right) The approximation $\hat{u}$ and $u$ for $a = 100$ showing very good fit. }
\label{fig:GoodnessAnecdote}
\end{figure}

\begin{figure}[htbp]
\centering
\includegraphics[width=0.65\textwidth]{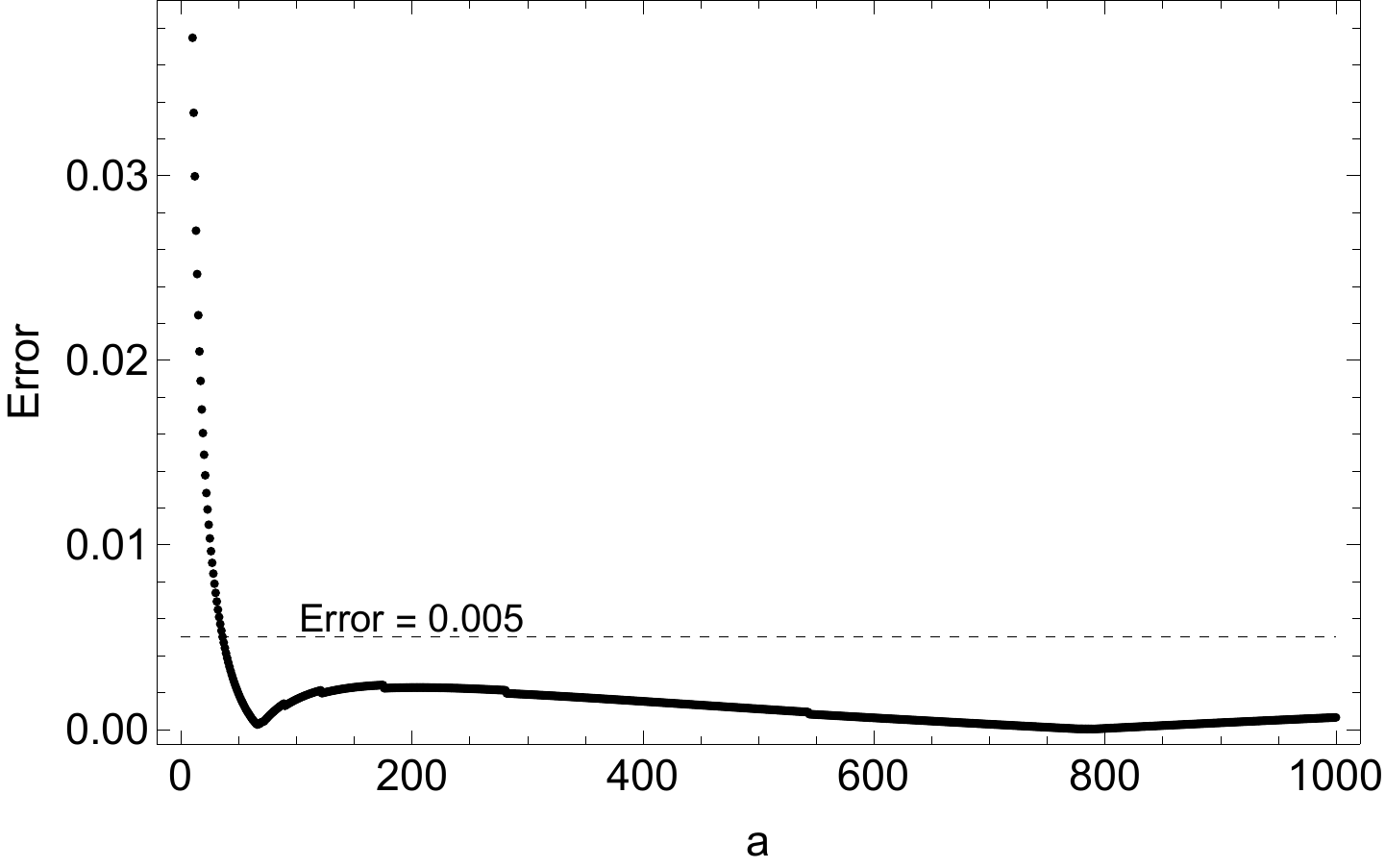}
\caption{The maximum error $E_\text{max}$ shows a rapid drop-off as $a$ increases. At $a = 40$, the error is well below $0.005$, where it remains.}
\label{fig:MaxError}
\end{figure}

Infinite domain problems are not always relevant to biological or physical problems. Consider the problem
\begin{equation}
\left\{
\begin{aligned}
\frac{\partial u}{\partial t} & = r u(1-u) + \frac{2k}{M}\frac{\partial M}{\partial x}\frac{\partial u}{\partial x} + k\frac{\partial^2 u}{\partial x^2}\\
\frac{\partial M}{\partial t} & = k\frac{\partial^2 M}{\partial x^2}\\
u(0) &= 1 \quad u(1) = 0.\\
M(0) &= b \quad M(1) = a+b\\
u(x,0) &= u_0(x) \quad M(x,0) = M_0(x).
\end{aligned}
\right.
\end{equation}
We assume $b,a+b > 0$. The population behaves according to a non-homogeneous heat equation, which has known solution. If we assume the population is at equilibrium so that $M(x) = ax + b$ and consider the equilibrium problem for $u(x,t)$, then the finite population Fisher-KPP equation with Dirichlet boundary conditions becomes,
\begin{equation}
\left\{
\begin{aligned}
&\frac{\partial u}{\partial t} = ru(1-u) + \frac{2ak}{ax+b}\frac{\partial u}{\partial x} + k\frac{\partial^2 u}{\partial x^2} \\
&u(0) = 1 \quad u(1) = 0.\\
&u(x,0) = u_0(x),
\end{aligned}
\right.
\end{equation}
which is the finite analogue of the infinite domain problem. Assume we define
\begin{equation*}
u_0(x) = \begin{cases}
1 & \text{if $x = 0$}\\
0 & \text{otherwise,}
\end{cases}
\end{equation*}
corresponding to an invasion of a stable spatially inhomogeneous population with $u(x) = 0$. The resulting dynamics for $a = 1$ and $a = 10$ and $k = r = b = 1$ are shown in \cref{fig:FiniteRegionSpaceTime}.
The figure illustrates distinct asymptotic behaviour for the two values of $a$ used. Assuming $k = r = 1$, the two-point boundary value problem describing the long-run behaviour is then
\begin{equation}
\begin{aligned}
&u'' + \frac{2a}{ax+b}u' + u(1-u) = 0\\
&u(0) = 1 \quad u(1) = 0.
\end{aligned}
\label{eqn:1DDirichletLaplace}
\end{equation}
This problem is not solvable by classical means, however the equations arising from linearization at the endpoints are both solvable in closed form. Suppose $u\approx 1$ and we replace the left-hand boundary condition with boundary conditions $u(0) = 1$ and $u'(0) = r_0 < 0$. The resulting linearized problem
\begin{equation*}
\begin{aligned}
&u'' + \frac{2a}{ax+b}u' + (1-u) = 0\\
&u(0) = 1 \quad u'(0) = r_L.
\end{aligned}
\end{equation*}
has closed form solution
\begin{equation*}
u_L(x;r_L) = \frac{b r_L \sinh (x)}{a x+b}+1.
\end{equation*}
The corresponding right-hand-side problem is given by
\begin{equation*}
\begin{aligned}
&u'' + \frac{2a}{ax+b}u' + u = 0\\
&u(1) = 0 \quad u'(1) = r_R,
\end{aligned}
\end{equation*}
with solution
\begin{equation*}
u_R(x;r_R) = -\frac{r_R (a+b) \sin (1-x)}{a x+b}.
\end{equation*}
As before, we know that $r_R \leq 0$.
Surprisingly, these solutions can be combined to construct an approximation for the solution of \cref{eqn:1DDirichletLaplace}, allowing us to explore the dynamics of the solutions. Let
\begin{equation*}
\hat{u}(x;q, r_L, r_R) = \begin{cases}
u_L(x;r_L) & \text{if $x \in [0,q)$}\\
u_R(x;r_R) & \text{if $x \in [q,1]$}.
\end{cases}
\label{eqn:Ansatz}
\end{equation*}
The point $q$ and values for $r_L$ and $r_R$ are not known and must be approximated for each input $(a,b)$. We illustrate this in \cref{fig:uhat}.
Assume we are given (a numerical solution for) $u(x)$. From this we can compute $u'(0) = r_L$ and $u'(1) = r_R$. It is easy to compute $q$ by solving the following optimization problem,
\begin{equation}
\arg\min_{q\in[0,1]}\;\; \int_0^1 \left[u(x) - \hat{u}(x;q, r_L, r_R)\right]^2\,dx,
\label{eqn:MinProblem}
\end{equation}
which can be accomplished using a simple dichotomous or golden section search \cite{BSS13}. Unfortunately, this still requires a numerical solution for $u(x)$ to determine $r_L$ and $r_R$. We can construct functions $r_L(a,b)$, $r_R(a,b)$ and $q(a,b)$ using a standard ordinary least squares approach. Without loss of generality, we fix $b = 1$ and show the resulting analysis by varying $a$ only. We use the following procedure:
\begin{enumerate}
\item Choose $a$ from a sample space $\mathcal{A}$ with $a \geq 0$.
\item Compute $u(x)$ using a numerical solver.
\item Compute $r_L = u'(0)$ and $r_R = u'(1)$ using the output of the numerical solver. 
\item Compute $q$ using \cref{eqn:MinProblem}.
\item Store $(a, q, r_L, r_R)$ for fitting. 
\end{enumerate}
Following this procedure leads to the data shown in \cref{fig:FitPlots}.  
The data can be fit with the following functions,
\begin{align}
\hat{q}(a) &\sim \frac{1}{(a + \alpha)^\beta},\\ 
\hat{r}_L(a) &\sim \alpha \cdot a^\beta, \text{ and}\\
\quad r_R(a) &\sim \frac{1}{\alpha \cdot a ^ \beta} .
\end{align}
Tables of fit values are given in the insets of \cref{fig:FitPlots}. The adjusted $r^2$ value for all three fits is $\sim0.9999$, suggesting a high degree of accuracy in the underlying model. Consequently, when $b=1$ an approximate solution to \cref{eqn:1DDirichletLaplace} is given in closed form by
\begin{equation*}
\hat{u}(x;a) \approx 
\begin{cases}
 \frac{\sinh (x) \left(-1.03 a^{0.996}\right)}{a x+1}+1 & 0\leq
   x<\frac{1}{(a+3.29)^{0.621}} \\
 -\frac{(a+1) \sin (1-x) \left(-0.883 a^{0.993}\right)}{a x+1} &
   \frac{1}{(a+3.29)^{0.621}}\leq x<1
\end{cases}
\end{equation*}
\cref{fig:GoodnessAnecdote} shows anecdotal evidence for the goodness of fit of this approximation for larger values of $a$. 
We compute the maximum error
\begin{equation*}
E_\text{max} = \max_{x}\left\lvert u(x;a) - \hat{u}(x;a)\right\rvert
\end{equation*}
for $a \in [0,1000]$. This is shown in \cref{fig:MaxError}. 
The figure shows that for $a > 40$, the maximum approximation error is well below $0.005$, where it remains as $a$ increases. When combined with the ansatz from \cref{eqn:Ansatz}, we can formulate the following conjecture (to be proved as future work). 
\begin{conjecture} Consider \cref{eqn:1DDirichletLaplace}. As $a \to \infty$, the solution $u$ approaches the weak solution
\begin{equation*}
u(x) = \begin{cases} 1 & \text{if x = 0}\\
0 & \text{if $0 < x \leq 1$}.
\end{cases}
\end{equation*}
\end{conjecture}
Interestingly, this behaviour is consistent with the behaviour observed in the travelling wave solution. As $a \to \infty$, the population at $x = 1$ becomes infinite. This is where $u(x) = 0$ . As the system comes to equilibrium, the infinite population overwhelms the finite population of invaders at $x = 0$, leading to the proposed weak solution. If we reversed the direction of population increase so that (e.g.) $M(x) = (a + 1) - ax$ and let $a \to \infty$, we would see a limiting population with $u(x) = 1$ for $0\leq x < 1$ and $u(1) = 0$.

\subsection{Neumann Boundary Conditions}
\begin{figure}[htbp]
\centering
\includegraphics[width=0.3\textwidth]{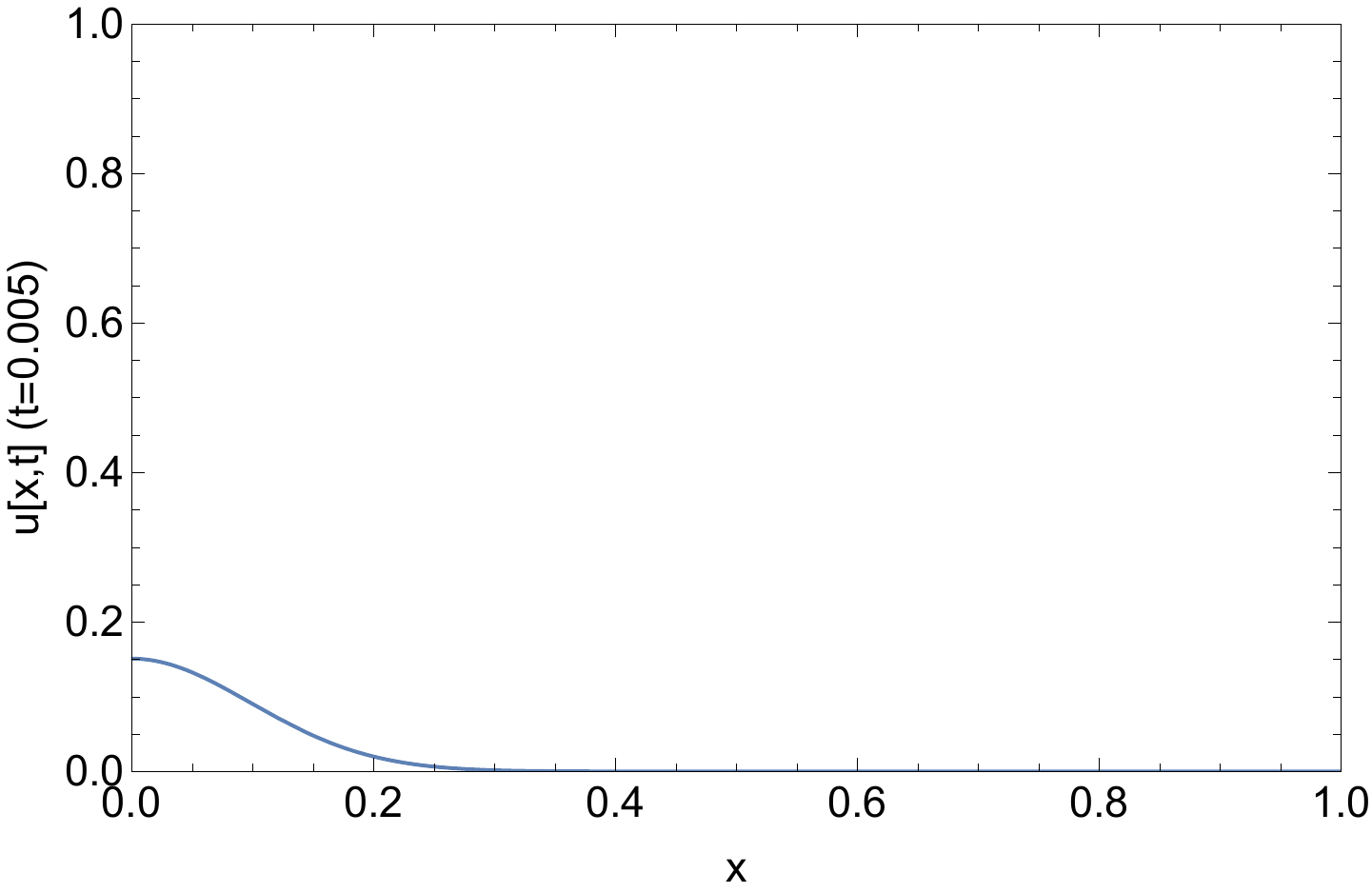}\quad
\includegraphics[width=0.3\textwidth]{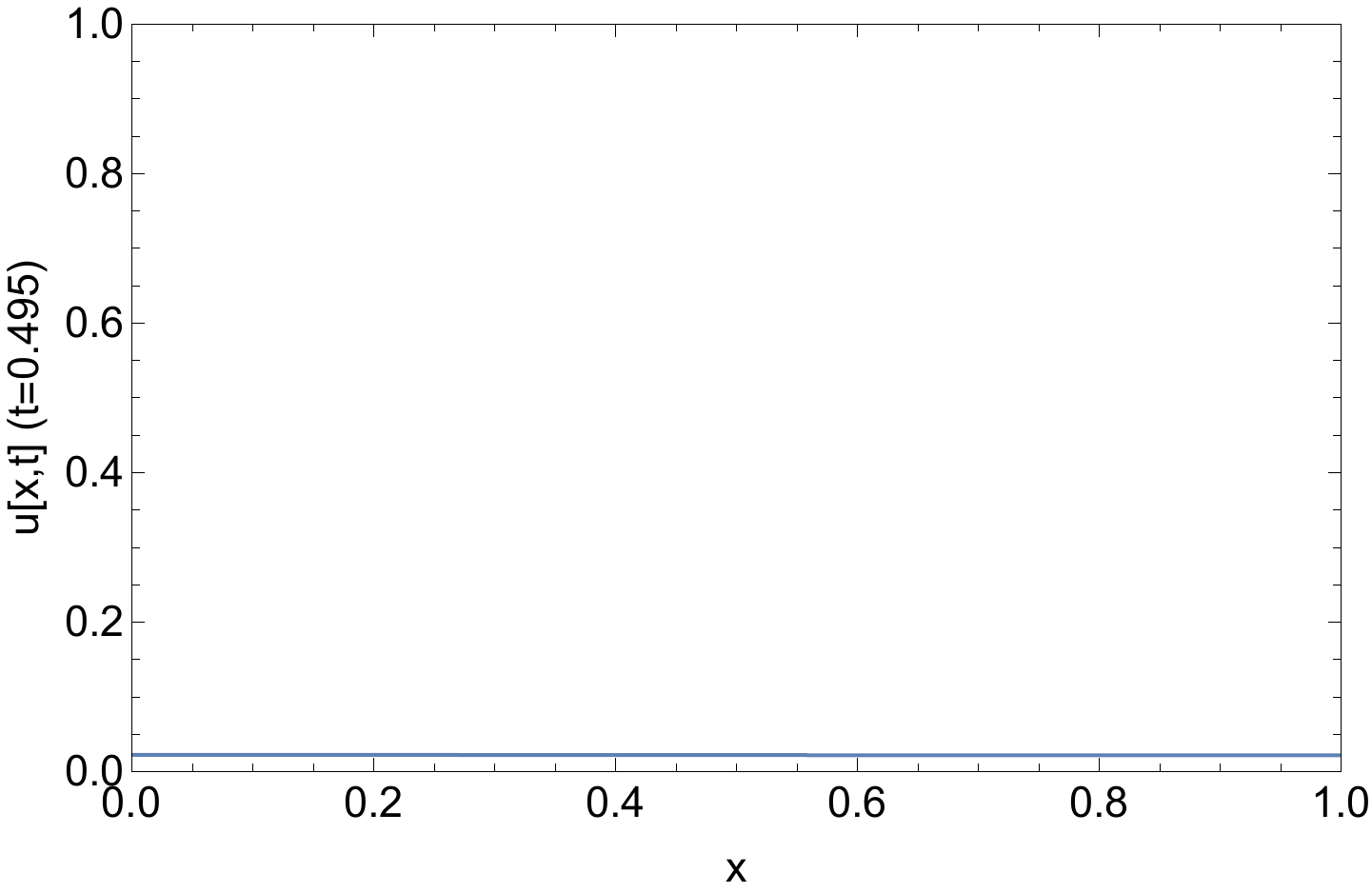}\quad
\includegraphics[width=0.3\textwidth]{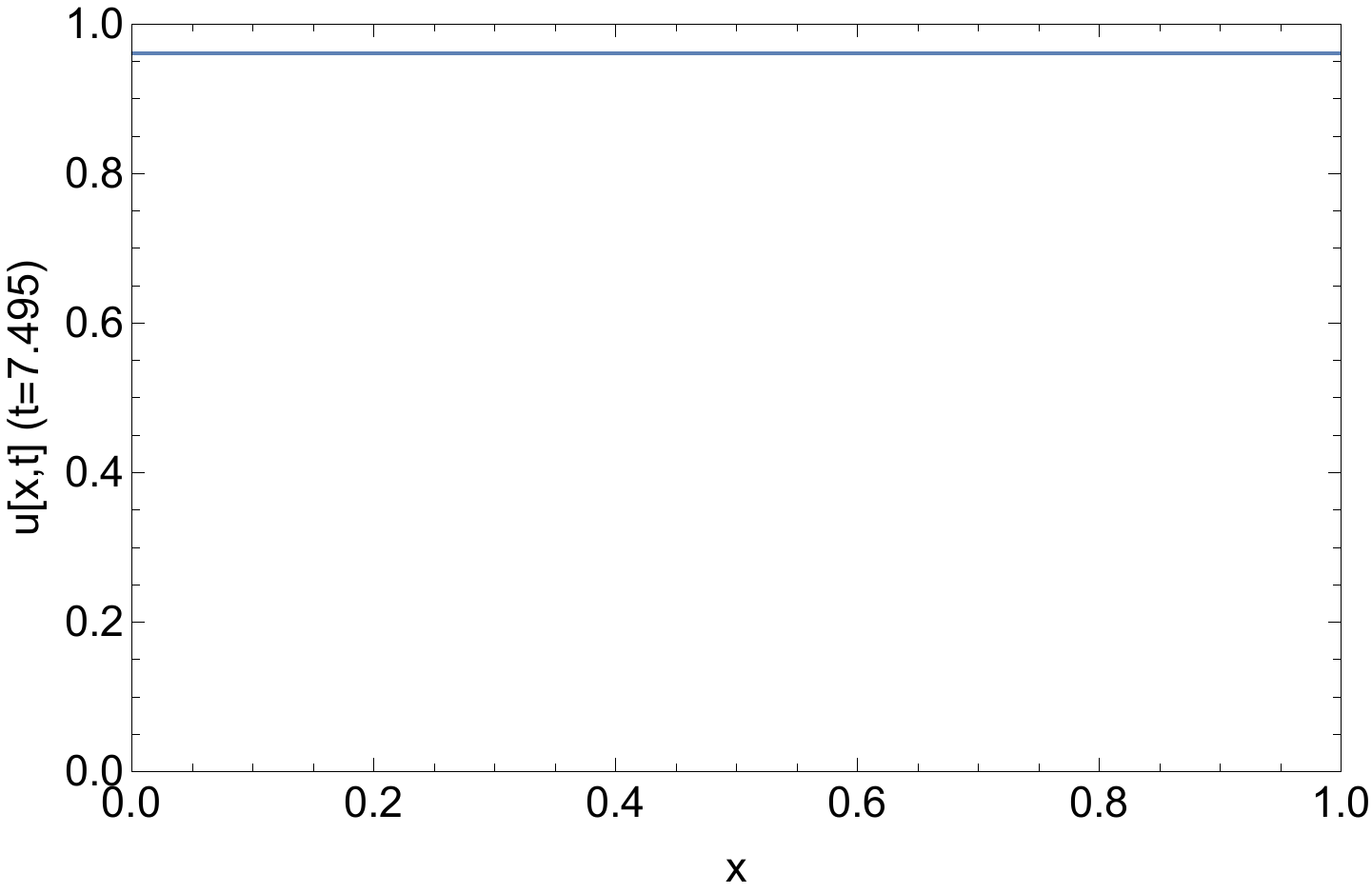}
\caption{The time evolution of $u(x,t)$ for $t = 0.005$ (left), $t = 0.495$ (middle) and $t = 7.495$ (right). This supports the conjecture that $\lim_{t\to\infty} u(x,t) = 1$.}
\label{fig:Neumann}
\end{figure}

It is worth briefly discussing \cref{eqn:1DDirichletLaplace} when the Dirichlet boundary conditions are replaced with Neumann boundary conditions.
\begin{equation}
\left\{
\begin{aligned}
&\frac{\partial u}{\partial t} = ru(1-u) + \frac{2ak}{ax+b}\frac{\partial u}{\partial x} + k\frac{\partial^2 u}{\partial x^2}  \\
&u_x(0) = 0 \quad u_x(1) = 0.\\
&u(x,0) = u_0(x),
\end{aligned}
\right.
\label{eqn:Neumann}
\end{equation}
The long-run behaviour of this system is relatively easy to predict.
\begin{conjecture} Let $u(x,t)$ be a solution to \cref{eqn:Neumann} with $u_0(x) > 0$ for some $x$. Then
\begin{equation*}
\lim_{t\to\infty} u(x,t) = 1.
\end{equation*}
\end{conjecture}
This conjecture is supported by numerical analysis. Fix $\epsilon > 0$ and let
\begin{equation*}
u_0(x) = 
\begin{cases} 
1 & \text{if $x < \epsilon$}\\
0 & \text{otherwise}.
\end{cases}
\end{equation*}
This initial condition is consistent with the Neumann boundary conditions. \cref{fig:Neumann} shows an early transient where $u(x,t)$ drops close to $0$ before the invasive species slowly takes over the entire population.
These dynamics are more consistent with the physical intuition associated with an invasive dominant species. 

\section{Conclusion}\label{sec:Conclusion}
In this paper, we studied the finite population Fisher-KPP equation, which arises naturally from the finite population spatial replicator using a skew-symmetric $2\times 2$ matrix. We showed using the results of Ablowitz and Zeppetella that this system of equations admits travelling wave solutions and showed a closed form solution that is identical to that in \cite{AZ79} except that the sign of the wave speed is reversed. We then studied the equilibrium problem for the finite population Fisher-KPP on a finite interval. We constructed a closed form approximate solution to this problem and used it to present a simple conjecture on the behaviour of populations with large spatial gradients.

There are several future directions that could be explored using the proposed finite population Fisher-KPP equation as a basis. For the purpose of this paper, we chose the travelling wave solution to the diffusion equation as the population equation. However, several closed form solutions to the diffusion equation exist and could be used, resulting in a new quasilinear reaction diffusion equation with a convection-like term. Also, in \cite{GMD21} it is noted there are no sensible stable amplitude travelling wave solutions when the logistic term $u(1-u)$ is replaced by a rock-paper-scissors dynamic, which arises from a $3 \times 3$ skew-symmetric payoff matrices. This paper shows that $2 \times 2$ skew-symmetric payoff matrices do give rise to travelling wave solutions in the finite population model. Therefore, it would be interesting to know whether any travelling wave solutions exist for $n\times n$ skew-symmetric payoff matrices with $n > 3$ or if this is purely a property of $2 \times 2$ payoff matrices. Proving the conjectures provided in this note are clearly a future direction of work with respect to the finite population Fisher-KPP equation in finite regions. However, an even more interesting direction might be to consider the problem on two-dimensional bounded regions (e.g., disks), where well known solutions to the Laplace equation are available. Studying more exotic boundary conditions (especially in two-dimensions) might also yield interesting results.

\section*{Acknowledgement}
Portions of C.G.'s work were supported by the National Science Foundation under grant CMMI-1932991.

\bibliography{FisherEquation}

\begin{thebibliography}{17}%
\makeatletter
\providecommand \@ifxundefined [1]{%
 \@ifx{#1\undefined}
}%
\providecommand \@ifnum [1]{%
 \ifnum #1\expandafter \@firstoftwo
 \else \expandafter \@secondoftwo
 \fi
}%
\providecommand \@ifx [1]{%
 \ifx #1\expandafter \@firstoftwo
 \else \expandafter \@secondoftwo
 \fi
}%
\providecommand \natexlab [1]{#1}%
\providecommand \enquote  [1]{``#1''}%
\providecommand \bibnamefont  [1]{#1}%
\providecommand \bibfnamefont [1]{#1}%
\providecommand \citenamefont [1]{#1}%
\providecommand \href@noop [0]{\@secondoftwo}%
\providecommand \href [0]{\begingroup \@sanitize@url \@href}%
\providecommand \@href[1]{\@@startlink{#1}\@@href}%
\providecommand \@@href[1]{\endgroup#1\@@endlink}%
\providecommand \@sanitize@url [0]{\catcode `\\12\catcode `\$12\catcode
  `\&12\catcode `\#12\catcode `\^12\catcode `\_12\catcode `\%12\relax}%
\providecommand \@@startlink[1]{}%
\providecommand \@@endlink[0]{}%
\providecommand \url  [0]{\begingroup\@sanitize@url \@url }%
\providecommand \@url [1]{\endgroup\@href {#1}{\urlprefix }}%
\providecommand \urlprefix  [0]{URL }%
\providecommand \Eprint [0]{\href }%
\providecommand \doibase [0]{https://doi.org/}%
\providecommand \selectlanguage [0]{\@gobble}%
\providecommand \bibinfo  [0]{\@secondoftwo}%
\providecommand \bibfield  [0]{\@secondoftwo}%
\providecommand \translation [1]{[#1]}%
\providecommand \BibitemOpen [0]{}%
\providecommand \bibitemStop [0]{}%
\providecommand \bibitemNoStop [0]{.\EOS\space}%
\providecommand \EOS [0]{\spacefactor3000\relax}%
\providecommand \BibitemShut  [1]{\csname bibitem#1\endcsname}%
\let\auto@bib@innerbib\@empty
%</preamble>
\bibitem [{\citenamefont {Fisher}(1937)}]{F37}%
  \BibitemOpen
  \bibfield  {author} {\bibinfo {author} {\bibfnamefont {R.~A.}\ \bibnamefont
  {Fisher}},\ }\bibfield  {title} {\bibinfo {title} {The wave of advance of
  advantageous genes},\ }\href@noop {} {\bibfield  {journal} {\bibinfo
  {journal} {Annals of eugenics}\ }\textbf {\bibinfo {volume} {7}},\ \bibinfo
  {pages} {355} (\bibinfo {year} {1937})}\BibitemShut {NoStop}%
\bibitem [{\citenamefont {Kolmogorov}\ \emph {et~al.}(1937)\citenamefont
  {Kolmogorov}, \citenamefont {Petrovskii},\ and\ \citenamefont
  {Piskunov}}]{KPP37}%
  \BibitemOpen
  \bibfield  {author} {\bibinfo {author} {\bibfnamefont {A.}~\bibnamefont
  {Kolmogorov}}, \bibinfo {author} {\bibfnamefont {I.}~\bibnamefont
  {Petrovskii}},\ and\ \bibinfo {author} {\bibfnamefont {N.}~\bibnamefont
  {Piskunov}},\ }\bibfield  {title} {\bibinfo {title} {A study of the diffusion
  equation with increase in the amount of substance, and its application to a
  biological problem in selected works of an kolmogorov, vol. 1, 242-270},\
  }\href@noop {} {\bibfield  {journal} {\bibinfo  {journal} {Bull. Moscow
  Univ., Math. Mech.}\ }\textbf {\bibinfo {volume} {1}},\ \bibinfo {pages} {1}
  (\bibinfo {year} {1937})}\BibitemShut {NoStop}%
\bibitem [{\citenamefont {Kenkre}(2004)}]{K04}%
  \BibitemOpen
  \bibfield  {author} {\bibinfo {author} {\bibfnamefont {V.}~\bibnamefont
  {Kenkre}},\ }\bibfield  {title} {\bibinfo {title} {Results from variants of
  the fisher equation in the study of epidemics and bacteria},\ }\href@noop {}
  {\bibfield  {journal} {\bibinfo  {journal} {Physica A: Statistical Mechanics
  and its Applications}\ }\textbf {\bibinfo {volume} {342}},\ \bibinfo {pages}
  {242} (\bibinfo {year} {2004})}\BibitemShut {NoStop}%
\bibitem [{\citenamefont {Sardar}\ \emph {et~al.}(2015)\citenamefont {Sardar},
  \citenamefont {Rana},\ and\ \citenamefont {Chattopadhyay}}]{SRC15}%
  \BibitemOpen
  \bibfield  {author} {\bibinfo {author} {\bibfnamefont {T.}~\bibnamefont
  {Sardar}}, \bibinfo {author} {\bibfnamefont {S.}~\bibnamefont {Rana}},\ and\
  \bibinfo {author} {\bibfnamefont {J.}~\bibnamefont {Chattopadhyay}},\
  }\bibfield  {title} {\bibinfo {title} {A mathematical model of dengue
  transmission with memory},\ }\href@noop {} {\bibfield  {journal} {\bibinfo
  {journal} {Communications in Nonlinear Science and Numerical Simulation}\
  }\textbf {\bibinfo {volume} {22}},\ \bibinfo {pages} {511} (\bibinfo {year}
  {2015})}\BibitemShut {NoStop}%
\bibitem [{\citenamefont {Beneduci}\ \emph {et~al.}(2021)\citenamefont
  {Beneduci}, \citenamefont {Bilotta},\ and\ \citenamefont {Pantano}}]{BBP21}%
  \BibitemOpen
  \bibfield  {author} {\bibinfo {author} {\bibfnamefont {R.}~\bibnamefont
  {Beneduci}}, \bibinfo {author} {\bibfnamefont {E.}~\bibnamefont {Bilotta}},\
  and\ \bibinfo {author} {\bibfnamefont {P.}~\bibnamefont {Pantano}},\
  }\bibfield  {title} {\bibinfo {title} {A unifying nonlinear probabilistic
  epidemic model in space and time},\ }\href@noop {} {\bibfield  {journal}
  {\bibinfo  {journal} {Scientific Reports}\ }\textbf {\bibinfo {volume}
  {11}},\ \bibinfo {pages} {1} (\bibinfo {year} {2021})}\BibitemShut {NoStop}%
\bibitem [{\citenamefont {Mehmet}(2012)}]{M12}%
  \BibitemOpen
  \bibfield  {author} {\bibinfo {author} {\bibfnamefont {M.}~\bibnamefont
  {Mehmet}},\ }\bibfield  {title} {\bibinfo {title} {Solutions of
  time-fractional reaction-diffusion equation with modified riemann-liouville
  derivative},\ }\href@noop {} {\bibfield  {journal} {\bibinfo  {journal}
  {International Journal of Physical Sciences}\ }\textbf {\bibinfo {volume}
  {7}},\ \bibinfo {pages} {2317} (\bibinfo {year} {2012})}\BibitemShut
  {NoStop}%
\bibitem [{\citenamefont {Neubert}\ and\ \citenamefont {Parker}(2004)}]{NP04}%
  \BibitemOpen
  \bibfield  {author} {\bibinfo {author} {\bibfnamefont {M.~G.}\ \bibnamefont
  {Neubert}}\ and\ \bibinfo {author} {\bibfnamefont {I.~M.}\ \bibnamefont
  {Parker}},\ }\bibfield  {title} {\bibinfo {title} {Projecting rates of spread
  for invasive species},\ }\href@noop {} {\bibfield  {journal} {\bibinfo
  {journal} {Risk Analysis: An International Journal}\ }\textbf {\bibinfo
  {volume} {24}},\ \bibinfo {pages} {817} (\bibinfo {year} {2004})}\BibitemShut
  {NoStop}%
\bibitem [{\citenamefont {Zel'Dovich}\ and\ \citenamefont
  {Raizer}(2002)}]{ZR02}%
  \BibitemOpen
  \bibfield  {author} {\bibinfo {author} {\bibfnamefont {Y.~B.}\ \bibnamefont
  {Zel'Dovich}}\ and\ \bibinfo {author} {\bibfnamefont {Y.~P.}\ \bibnamefont
  {Raizer}},\ }\href@noop {} {\emph {\bibinfo {title} {Physics of shock waves
  and high-temperature hydrodynamic phenomena}}}\ (\bibinfo  {publisher}
  {Courier Corporation},\ \bibinfo {year} {2002})\BibitemShut {NoStop}%
\bibitem [{\citenamefont {Kametaka}(1976)}]{K76}%
  \BibitemOpen
  \bibfield  {author} {\bibinfo {author} {\bibfnamefont {Y.}~\bibnamefont
  {Kametaka}},\ }\bibfield  {title} {\bibinfo {title} {On the nonlinear
  diffusion equation of kolmogorov-petrovskii-piskunov type},\ }\href@noop {}
  {\bibfield  {journal} {\bibinfo  {journal} {Osaka Journal of Mathematics}\
  }\textbf {\bibinfo {volume} {13}},\ \bibinfo {pages} {11} (\bibinfo {year}
  {1976})}\BibitemShut {NoStop}%
\bibitem [{\citenamefont {Uchiyama}(1977)}]{U77}%
  \BibitemOpen
  \bibfield  {author} {\bibinfo {author} {\bibfnamefont {K.}~\bibnamefont
  {Uchiyama}},\ }\bibfield  {title} {\bibinfo {title} {The behavior of
  solutions of the equation of kolmogorov-petrovsky-piskunov},\ }\href@noop {}
  {\bibfield  {journal} {\bibinfo  {journal} {Proceedings of the Japan Academy,
  Series A, Mathematical Sciences}\ }\textbf {\bibinfo {volume} {53}},\
  \bibinfo {pages} {225} (\bibinfo {year} {1977})}\BibitemShut {NoStop}%
\bibitem [{\citenamefont {Newman}(1980)}]{N80}%
  \BibitemOpen
  \bibfield  {author} {\bibinfo {author} {\bibfnamefont {W.~I.}\ \bibnamefont
  {Newman}},\ }\bibfield  {title} {\bibinfo {title} {Some exact solutions to a
  non-linear diffusion problem in population genetics and combustion},\
  }\href@noop {} {\bibfield  {journal} {\bibinfo  {journal} {Journal of
  Theoretical Biology}\ }\textbf {\bibinfo {volume} {85}},\ \bibinfo {pages}
  {325} (\bibinfo {year} {1980})}\BibitemShut {NoStop}%
\bibitem [{\citenamefont {Weinberger}(1982)}]{W82}%
  \BibitemOpen
  \bibfield  {author} {\bibinfo {author} {\bibfnamefont {H.~F.}\ \bibnamefont
  {Weinberger}},\ }\bibfield  {title} {\bibinfo {title} {Long-time behavior of
  a class of biological models},\ }\href@noop {} {\bibfield  {journal}
  {\bibinfo  {journal} {SIAM journal on Mathematical Analysis}\ }\textbf
  {\bibinfo {volume} {13}},\ \bibinfo {pages} {353} (\bibinfo {year}
  {1982})}\BibitemShut {NoStop}%
\bibitem [{\citenamefont {Ablowitz}\ and\ \citenamefont
  {Zeppetella}(1979)}]{AZ79}%
  \BibitemOpen
  \bibfield  {author} {\bibinfo {author} {\bibfnamefont {M.~J.}\ \bibnamefont
  {Ablowitz}}\ and\ \bibinfo {author} {\bibfnamefont {A.}~\bibnamefont
  {Zeppetella}},\ }\bibfield  {title} {\bibinfo {title} {Explicit solutions of
  fisher's equation for a special wave speed},\ }\href@noop {} {\bibfield
  {journal} {\bibinfo  {journal} {Bulletin of Mathematical Biology}\ }\textbf
  {\bibinfo {volume} {41}},\ \bibinfo {pages} {835} (\bibinfo {year}
  {1979})}\BibitemShut {NoStop}%
\bibitem [{\citenamefont {Griffin}\ \emph {et~al.}(2021)\citenamefont
  {Griffin}, \citenamefont {Mummah},\ and\ \citenamefont {DeForest}}]{GMD21}%
  \BibitemOpen
  \bibfield  {author} {\bibinfo {author} {\bibfnamefont {C.}~\bibnamefont
  {Griffin}}, \bibinfo {author} {\bibfnamefont {R.}~\bibnamefont {Mummah}},\
  and\ \bibinfo {author} {\bibfnamefont {R.}~\bibnamefont {DeForest}},\
  }\bibfield  {title} {\bibinfo {title} {A finite population destroys a
  traveling wave in spatial replicator dynamics},\ }\href@noop {} {\bibfield
  {journal} {\bibinfo  {journal} {Chaos, Solitons \& Fractals}\ }\textbf
  {\bibinfo {volume} {146}},\ \bibinfo {pages} {110847} (\bibinfo {year}
  {2021})}\BibitemShut {NoStop}%
\bibitem [{\citenamefont {Vickers}(1991)}]{V91}%
  \BibitemOpen
  \bibfield  {author} {\bibinfo {author} {\bibfnamefont {G.}~\bibnamefont
  {Vickers}},\ }\bibfield  {title} {\bibinfo {title} {Spatial patterns and
  travelling waves in population genetics},\ }\href@noop {} {\bibfield
  {journal} {\bibinfo  {journal} {Journal of Theoretical Biology}\ }\textbf
  {\bibinfo {volume} {150}},\ \bibinfo {pages} {329} (\bibinfo {year}
  {1991})}\BibitemShut {NoStop}%
\bibitem [{\citenamefont {Durrett}\ and\ \citenamefont {Levin}(1994)}]{DL94}%
  \BibitemOpen
  \bibfield  {author} {\bibinfo {author} {\bibfnamefont {R.}~\bibnamefont
  {Durrett}}\ and\ \bibinfo {author} {\bibfnamefont {S.}~\bibnamefont
  {Levin}},\ }\bibfield  {title} {\bibinfo {title} {{The Importance of Being
  Discrete (and Spatial)}},\ }\href@noop {} {\bibfield  {journal} {\bibinfo
  {journal} {Theoretical Population Biology}\ }\textbf {\bibinfo {volume}
  {46}},\ \bibinfo {pages} {363} (\bibinfo {year} {1994})}\BibitemShut
  {NoStop}%
\bibitem [{\citenamefont {Bazaraa}\ \emph {et~al.}(2013)\citenamefont
  {Bazaraa}, \citenamefont {Sherali},\ and\ \citenamefont {Shetty}}]{BSS13}%
  \BibitemOpen
  \bibfield  {author} {\bibinfo {author} {\bibfnamefont {M.~S.}\ \bibnamefont
  {Bazaraa}}, \bibinfo {author} {\bibfnamefont {H.~D.}\ \bibnamefont
  {Sherali}},\ and\ \bibinfo {author} {\bibfnamefont {C.~M.}\ \bibnamefont
  {Shetty}},\ }\href@noop {} {\emph {\bibinfo {title} {Nonlinear programming:
  theory and algorithms}}}\ (\bibinfo  {publisher} {John Wiley \& Sons},\
  \bibinfo {year} {2013})\BibitemShut {NoStop}%
\end{thebibliography}%
\end{document}